\documentclass[10pt,conference]{IEEEtran}

\IEEEoverridecommandlockouts

\usepackage{amsmath,amsfonts}
\usepackage[ruled,vlined]{algorithm2e}

\usepackage{graphicx}
\usepackage{array}
\usepackage{amssymb}
\usepackage{amsthm}
\usepackage{graphicx}
\usepackage{color}
\usepackage[style=ieee, backend=biber]{biblatex}

\usepackage{tabularx}

\newtheorem{thm}{Lemma}[section]

\begin{document}
\title{Motivating Learners in Multi-Orchestrator Mobile Edge Learning: A Stackelberg Game Approach}
\author{Mhd Saria Allahham$^1$, Sameh Sorour$^1$,  Amr Mohamed$^2$, Aiman Erbad$^3$ and Mohsen Guizani$^2$\\
		$^1$School of Computing, Queen's University, Kingston, ON, Canada
		\\
		$^2$College of Engineering, Qatar University, Qatar \\
		$^3$College of Science and Engineering, Hamad Bin Khalifa University, Qatar\\ 
		Email: \{20msa7, sameh.sorour\}@queensu.ca, amrm@qu.edu.qa, and \{aerbad, mguizani\}@ieee.org 
}

\maketitle

\begin{abstract}
Mobile Edge Learning (MEL) is a learning paradigm that enables distributed training of Machine Learning models over heterogeneous edge devices (e.g., IoT devices). Multi-orchestrator MEL refers to the coexistence of multiple learning tasks with different datasets, each of which being governed by an orchestrator to facilitate the distributed training process. In MEL, the training performance deteriorates without the availability of sufficient training data or computing resources. Therefore, it is crucial to motivate edge devices to become learners and offer their computing resources, and either offer their private data or receive the needed data from the orchestrator and participate in the training process of a learning task. In this work, we propose an incentive mechanism, where we formulate the orchestrators-learners interactions as a 2-round Stackelberg game to motivate the participation of the learners. In the first round, the learners decide which learning task to get engaged in, and then in the second round, the training parameters and the amount of data for training in case of participation such that their utility is maximized. We then study the training round analytically and derive the learners' optimal strategy. Finally, numerical experiments have been conducted to evaluate the performance of the proposed incentive mechanism.

\end{abstract}

\section{Introduction}
The unprecedented availability of edge nodes with the huge amount of its generated data has opened the doors for Machine Learning (ML)-driven services at the edge such as surveillance \cite{edge_surveillance} and crowdsensing \cite{edge_crowd}. Each of these services requires the training of ML models on the collected data in order to be deployed. Cloud-based schemes for training the models require the transfer of the data from a massive number of edge nodes to the cloud, which can be inefficient in terms of resources usage and causes a huge latency overhead. 
Alternatively, edge-based schemes have proven to be more efficient, where the training of ML models can be done in a distributed manner by the utilization of the available computing resources at the edge \cite{gunduz2020communicate}. 
Thanks to the advancements in Mobile Edge Computing (MEC) \cite{edge2017survey} and distributed ML \cite{mcmahan2017communication, edge_control}, a new paradigm for distributed training over the edge devices named Mobile Edge Learning (MEL) has been proposed in \cite{ME_1st}.
MEL is a cooperative paradigm that efficiently distributes and executes the learning tasks (i.e., ML models’ training) on heterogeneous wireless edge nodes such as IoT devices.  
The two main components of MEL are 1) orchestrator, which distributes the learning tasks synchronize the updates of the tasks, and 2) learners, which are capable of and responsible for training the local ML models on their own possessed or received data. MEL is a generalization of both Federated learning (FL) and Parallelized Learning (PL). In the former, the data is available at the learners' side, where the aim is to learn a global model across all the learners without the exposure of the private data. Whereas in the latter, the data is at the orchestrator side, but it lacks the computation capabilities to perform the learning task by itself. Thus, it distributes the task and the data across trusted learners to utilize their resources. However, learners are reluctant to get engaged in the learning process without receiving benefits or compensation, either for offering their private data to train on, or for offering their computing resources and receive data from other incapable nodes to help in performing the learning task. In either case, the learning experience for the whole task increases with the amount and quality of data included in training \cite{zhan2020learning_incentive}. As such, it is essential to motivate as many learners as possible to participate in the learning process in an active and reliable manner.

The related work in the literature exploits different methodologies for developing incentive mechanisms for the learners including Game Theory \cite{zhan2020learning_incentive, sarikaya2019motivating_SB1} and Auction Theory \cite{zeng2020fmore_auction1, le2021incentive_auction2}. For example, in \cite{sarikaya2019motivating_SB1}, the interactions between the learners and the orchestrator have been modeled as a Stackelberg game, where the learners try to determine the computation power that they will use in the training based on the incentive provided by the orchestrator. Whereas the latter tries to minimize the total training time to achieve a certain learning accuracy and optimize the budget allocation to the learners. A similar Stackelberg game formulation has been proposed in \cite{zhan2020learning_incentive}, but instead, the learners try to determine the amount of data that will be included in the training, and the orchestrator aims to maximize the learning experience.
As for auction-based incentives, the authors in \cite{zeng2020fmore_auction1} developed a generic lightweight algorithm for encouraging learners' participation, taking into account the learner's multiple resources. The top $K$ learners with the highest bids are then selected to be engaged in the learning process. Whereas in \cite{le2021incentive_auction2}, the authors formulated the incentive mechanism with wireless cellular network characteristics. The learners decide their bid and policy in terms of transmission power, computation capacity based on an allocated bandwidth and incentive. The learner selection is then formulated as a knapsack problem, and solved via a greedy-based algorithm.

However, the previous works do not consider the heterogeneity of the learners' capabilities and its effect on the learning process. Moreover, they have focused on developing incentive mechanisms assuming a single orchestrator and single learning task in the system, and none of them considered the multi-orchestrator case. In multi-orchestrator MEL, each orchestrator has a different independent learning task. Accordingly, the learners get to choose which learning task to get engaged in (i.e., which orchestrator to associate with) such that the service costs are minimized and the utility is maximized. To this end, we aim to design an incentive mechanism for the multi-orchestrator MEL system, where learners first decide the association, and then decide the amount of data that they will train on in the learning process based on the associated orchestrator's incentive, while the latter decides the incentive and the training parameters based on the learners' capabilities. The main contributions of the paper can be summarized as follows:

\begin{enumerate}
    \item First, we formulate the learners-orchestrators interactions as a single 2-round Stackelberg game, where the associations are decided in the first round, and the training parameters and the incentives are decided in the second round. 
    \item In the first round, as the association problem turned out to be NP-hard, we propose a heuristic approach for the learners-orchestrators association. Whereas in the second round game, we prove the existence of a Nash equilibrium, where we derive the optimal strategies for both learners and orchestrators.
    \item Finally, we carry out numerical results to show the performance of the proposed incentive mechanism while being compared to other techniques.
\end{enumerate}

The remainder of the paper is organized as follows: We introduce the system model in Section \ref{sec:sys_model}. Section \ref{sec:form} describes the formulated Stackelberg game, including the orchestrators' and learners' utility functions, while Section \ref{sec:sol} presents the solution approach. We show the simulation results in Section \ref{sec:sim} before we conclude in Section \ref{sec:conc}.

\section{System Model}\label{sec:sys_model}
In this paper, we consider a multi-orchestrator multi-task MEL system. Each of these orchestrators can be either 1) a governing node for learners that have private data for the same learning task in the case of FL, or 2) an edge node that is incapable of doing the training of its learning problem due to the lack of its computing resources in the case of PL. Nevertheless, our formulations are generic and applicable for both cases, but we will focus on the PL case in this work and point out the difference in the formulation whenever needed. 

\subsection{Learning Settings}
We denote the set of learners by $\mathcal{L}$ and the set of learners that are associated with orchestrator $o$ by  $\mathcal{L}_o$, and each orchestrator has a dataset with $N_o$ samples. After the association, an orchestrator $o$ sends to a participating learner $l$ the learning task in terms of the model parameters $\boldsymbol{w}_{l,o}$, and $n_{l,o}$ data samples to train on \footnote{In case of FL, $n_{l,o}$ represents the number of data samples from the local dataset of the learner.}.
Afterward, each participating learner $l$ performs $\tau_{l,o}$ local training iterations employing Stochastic Gradient Descent (SGD) to minimize its local loss function. Once done, the learners send back to the orchestrator the training model parameters, where the latter aggregates these parameters by performing weighted averaging on the received models.

Each orchestrator then keeps sending back the data and the updated model for $G_o$ global cycles until a stopping criterion is satisfied such as the exhaustion of the resources, e.g., energy. Moreover, we consider the system is globally synchronous and locally asynchronous as in \cite{mohammad2020task_asynchronous}. In other words, at each global cycle, all the models from all the learners are collected and aggregated, but between the global cycles, each learner performs a different number of local training iterations on their local models. By considering so, the effect of the "straggler’s dilemma" is reduced, which represents how the learning process is throttled by the learner with the least capabilities \cite{cai2020d2d_straggler}.
Last but not least, we adopt the presented model in \cite{allahham2021energy} to define the learning objective as a function of the local iterations and global cycles as follows:
\begin{equation}\label{eqn:learning}
\tilde{F}_o(\tau_{l,o}, G_o) =  \frac{c1}{G_o \tau^{c2}_{l,o}}
\end{equation}
where $\tau_{l,o} \in [1,\tau_{max}]$ to ensure the convexity of the objective, c1 and c2 are constants that depend on learning parameters, namely, the divergence between the learners' local models and the global model, and the learning rates. The objective function (\ref{eqn:learning}) represents the distributed learning convergence bounds over
the edge network. The convergence bounds refer to how much the trained global model in the distributed learning is deviating from the optimal model. Therein, we denote the distributed learning quality as $F_o(\tau_{l,o}, G_o) = - \tilde{F}_o(\tau_{l,o}, G_o)$. 
Since the objective function is obviously convex for $c1,c2> 0$, it follows that the learning quality $F_o(\tau_{l,o}, G_o)$ is a concave function. 
Interested readers can refer to \cite{edge_control} and \cite{allahham2021energy} for more details about the learning objective formulation.

\subsection{Mobile Edge Settings}
We introduce the communication and computation models for wireless edge learning in this part. First, the number of bits that a learner $l$ receives from an orchestrator $o$ can be defined as:
\begin{equation}
B^{data}_{l,o} = n_{l,o} X_o\Gamma^d_o
\end{equation}
\begin{equation}
B^{weights}_{o} = S^w_o \Gamma^w_o
\end{equation}
where $X_o$ is the feature vector length, $S^w_o$ the total number of weights in the model, and $\Gamma^d_o$ and $\Gamma^w_o$ represent the bits/feature and bits/weight values, respectively. Consequently, the transmission time needed for both the data and the model weights from the orchestrator, and the updated model weights from the learner can be given by,  respectively\footnote{In case of FL, the term $B^{data}_{l,o}$ is set to 0.}:
\begin{equation}
t^S_{l,o} =  \frac{B^{data}_{l,o}+ B_o^{weights}}{W \log_2 (1+ \frac{h_{l,o} P}{\sigma^2})} 
\end{equation}
\begin{equation}
t^U_{l,o} =  \frac{B^{weights}_{o}}{W \log_2 (1+ \frac{h_{l,o} P}{\sigma^2})}
\end{equation}
where $P$ is the devices' transmission power, $W$ is the channel bandwidth, $\sigma^2$ is the channel noise variance, and $h_{l,o}$ is the channel gain expressed as $h_{l,o} = d^{-\nu}_{l,o} g^2$ where $\nu$ is the path loss exponent, $d_{l,o}$ is the distance between the orchestrator and the learner, and $g$ is the fading channel coefficient. It is worth noting that we assume the bandwidth and the transmission power are fixed during the training, as well as channel reciprocity. We then define learner's training time as follows:
\begin{equation}
t^C_{l,o} = \frac{\tau_{l,o} n_{l,o} C_o}{f_l}
\end{equation}
where $f_l$ is the learner's CPU frequency and $C_o$ is the model computational complexity (i.e., the number of CPU cycles needed to train on a single data sample).

Subsequently, we define the energy consumption models in the system. As the energy consumed for communications is the product of the transmission power with the transmission time, a learner's communication energy consumption can be given by: 
\begin{equation}
E^U_{l,o} = P t^U_{l,o} = \frac{P_{l,o} B^{weights}_{o}}{W \log_2 (1+ \frac{h_{l,o} P_{l,o}}{\sigma^2})}
\end{equation}
We also denote the reception energy consumption as $E^S_{l,o}$, which we consider it as a constant and is coupled with the number of data samples $n_{l,o}$ to be received from the orchestrator\footnote{In case of FL, the term $E^S_{l,o}$ is set to 0.}. As for the learner's computation energy consumption, we adopt the energy model in \cite{mao2017survey_edge_energy}, which can be defined in our system as:

\begin{equation}
E^C_{l,o} =\frac{\mu \tau_{l,o} n_{l,o} C_o}{f^{\xi}_l} 
\end{equation}
where $\mu$ and $\xi$ are hardware-related constants. 

\section{Incentive Mechanism Formulation}\label{sec:form}
We formulate the incentive mechanism for multi-orchestrator MEL as a static 2-rounds Stackelberg game \cite{myerson2013game} for the whole learning process, where the players are the orchestrators and the learners. There are two stages in each round, where the orchestrators are the leaders which act in the first stage, and the learners are the followers which act in the second stage. 

In the first round, the orchestrators first decide the initial payment and broadcast their learning task details including: 1) the time frame of the whole learning process $T_{max}$, 2) the initial monetary service price $\rho^{i}_o$ (i.e., \$/CPU cycle), 3) the minimum number of local iterations $\tau_{min}$ and global cycles $G_{min}$ to guarantee a learning experience for the task. The learners then choose which orchestrator to associate with, and send their computing and communication capabilities information. In the second round, each orchestrator $o$ then announces its strategy including the final monetary service price for each learner $\rho_{l,o}$ in the first stage, followed by the second stage where each learner determines its strategy in terms of the number of data samples $n_{l,o}$ such that their utility is maximized. We express an orchestrator's strategy in the second round as the number of local iterations to be performed by each associated learner and the monetary service price. Consequently, we can define the utility of the orchestrator $o$ as:
\begin{equation}
U_o(\tau_{l,o}, \rho_{l,o}) = \sum_l \lambda_{l,o}   \left  ( F_o(\tau_{l,o}, G_o) - G_o C_o \rho_{l,o} \tau_{l,o} n_{l,o}  \right )
\end{equation}
where the $\lambda_{l,o}$ is the association variable, and the first term represents the task learning quality, while the second term represents the total payment to the learners. However, the orchestrators cannot determine the number of global cycles since the considered MEL system is globally synchronous, and the learners have heterogeneous capabilities. Hence, the learning time will be dependent on the weakest learner that causes the maximum computation and communication delay. As a result, the orchestrator has to set the number of global cycles according to that learner to accommodate all the learners in the training process and increase the level of participation. In fact, by considering the weakest learner, the number of global cycles will be limited, but the orchestrator will have more degrees of freedom in determining its strategy in terms of the monetary service price and the number of local iterations for the other learners. After determining the number of global cycles, an orchestrator needs to determine its strategy by maximizing its utility, i.e., $\underset{\tau_{l,o}, \rho_o}{\max.}~ U_o(\tau_{l,o}, \rho_{l,o})$.
As for the learners, their strategy is expressed as the association with an orchestrator, which will be done the first round, and the amount of participation if decided in terms of the data samples in the second round. Moreover, we assume the distribution of the decided training data is representative of the whole data distribution, and the data at each learner have the same quality, i.e., independently and identically distributed.
We define the service cost for the learners as their computation and communication energy consumption, which are proportional to the amount of training data. Therein, we can define a learner $l$ utility as follows:
\begin{equation}
U_l(\lambda_{l,o}, n_{l,o}) = \sum_o \lambda_{l,o} G_o \left   (  C_o \rho_{l,o} \tau_{l,o} n_{l,o} - ( E^S_{l,o} + E^U_{l,o} + E^C_{l,o} )\right )
\end{equation}
where the first term represents the total revenue from the learning process. We assume that a learner can associate with only one orchestrator for a single learning task, and it has to finish the training within the learning task time frame.
Accordingly, the learners can decide their optimal strategy by solving the following optimization problem:
\begin{subequations}\label{eqn:opt_P1}
\begin{align}
& \mathbf{P1}: \;\;\; \underset{\lambda_{l,o}, n_{l,o}}{\max.} U_l(\lambda_{l,o}, n_{l,o})\\
&s.t. \sum_o \lambda_{l,o} \left (t^S_{l,o} + t^U_{l,o} + t^C_{l,o} \right ) \leq T_{max} \\
& \quad\;\; \sum_o \lambda_{l,o} = 1 \\
&\quad\;\;  0\leq n_{l,o} \leq n_{max} \\
&\quad \;\; \lambda_{l,o} \in \{ 0, 1\}
\end{align}
\end{subequations}
where $T_{max}$ is the learning time period, and $n_{max}$ is a hyperparameter that indicates the maximum allowed number of data samples to receive from the orchestrator if any, where $n_{max} \in (0,1]$. Lastly, to simplify the formulations in the reminder of the paper, we define the following coefficients: 
\begin{description}\small
        \item $A^0_{l,o} = \frac{2  B^{weights}_{o}}{T_{max} \times W \log_2 (1+ \frac{h_{l,o} P_{l,o}}{\sigma^2})}$, $\;\;\;\;\zeta^0_{l,o} = \frac{P_{l,o} A^0_{l,o}}{E_{max}}$
        \item $\\$
        \item $A^1_{l,o} = \frac{N_o F_o \Gamma_o^d}{T_{max}\times W \log_2 (1+ \frac{h_{l,o} P_{l,o}}{\sigma^2})}$, $\;\;\;\;\zeta^1_{l,o}= E^S_{l,o}$
        \item $\\$
        \item $A^2_{l,o} =  \frac{N_o C^w_o}{T_{max} \times f_l}$, $\;\;\;\;\;\;\;\;\;\;\;\;\;\;\;\;\;\;\;\zeta^2_{l,o}= (\frac{C_o \rho_{l,o}}{R_{max}} - \frac{\mu C_o}{ f^\xi_lE_{max}})$
    \end{description}
where $E_{max}$ and $R_{max}$ are the maximum energy consumption and revenue for the learners, respectively.


\section{Solution Approach}\label{sec:sol}
In this section, we first present a heuristic approach for the learners-orchestrator association problem. Afterward, we analyze the learners' and orchestrators' behavior and derive the optimal learner strategy in the second round.

\subsection{First round: Factor-Based Association}
Since the learners in the first round determine the association given the learning task details, the optimization problem \textbf{P1} can be first solved to determine the association by assuming a fixed arbitrary $n_{l,o}$ for all the learners. Herein, we define the association sub-problem \textbf{SP1} as follows:
\begin{subequations}\label{eqn:reform_opt}
\begin{align}
&\mathbf{SP1}: \;\;\; \underset{\lambda_{l,o}}{\max.}~~~  \sum_o \lambda_o G_o \left ( n_{l,o} ( \zeta^2_{l,o}\tau_{l,o} - \zeta^1_{l,o}) - \zeta^0_{l,o} \right )\\
&s.t.\;\;\;\,\sum \lambda_o G_o (A^2_{l,o} \tau_{l,o} n_{l,o} + A^1_{l,o} n_{l,o} + A^0_{l,o}  ) \leq 1 \\
&\quad\;\;\;\;\;\sum_o \lambda_{l,o} = 1 \\
&\quad\;\;\;\;\;\lambda_o \in \{0,1\}
\end{align}
\end{subequations}

It can be noticed that \textbf{SP1} is a binary integer linear program (BILP), which is eventually a minimization knapsack problem, and is known to be NP-hard. As a result, employing algorithms to derive an equilibrium based on estimating the learners strategy is not feasible in the first round.
Thus, we first present a heuristic algorithm for learners-orchestrators association that we refer to as the Factor-Based Association (FBA). We define the association factor (AF) with learner $l$ and orchestrator $o$ as follows:
\begin{equation}
\Gamma_{l,o} = \frac{\tilde{\rho}^{i}_o \tilde{C_o}}{\tilde{d}_{l,o}}
\end{equation}
where the term $\tilde{\rho}^{i}_o \tilde{C_o}$ represents the normalized initial payment per data sample from the orchestrator , and $\tilde{d}_{l,o}$ is the normalized distance between learner $l$ and orchestrator $o$, respectively, and both $\in [0,1]$. The AF is calculated at the learners' side in order for them to decide the association with which orchestrator. The AF characterizes the potential revenue for the learner, and since the game is played only once and does not capture the communication channel dynamics, the AF also characterizes the connectivity by the distance. Each learner then associate with orchestrator $\tilde{o}$ such that $\tilde{o} = \arg \underset{o}{\max} ~ \Gamma_{l,o}$. After the association is done, each learner sends to its associated orchestrator information about its computation and communication capabilities. 

\subsection{Second round: Deriving Learners' Strategy}
In the second round, we employ the backward-induction method to derive
the Stackelberg equilibrium, where the second stage in this round is solved to obtain the learners' optimal strategy, which is then used for solving the first stage to obtain the associated orchestrator's optimal strategy. Since the variable $n_{l,o}$ is an integer, we relax it to a continuous variable to solve the problem, then we round it to the nearest integer in the solution.  

For a given association $\lambda_{l,o}$, monetary service price $\rho_{l,o}$, local training iterations $\tau_{l,o}$ and global cycles $G_o$, the learner can determine its strategy by solving the participation sub-problem \textbf{SP2} which can be given as the following :
\begin{subequations}\label{eqn:reform_opt}
\begin{align}
&\mathbf{SP2}: \;\;\; \underset{n_{l,o}}{\max.}~~~   n_{l,o} ( \zeta^2_{l,o}\tau_{l,o} - \zeta^1_{l,o}) \\
&s.t.\;\;\; G_o (A^2_{l,o} \tau_{l,o} n_{l,o} + A^1_{l,o} n_{l,o} + A^0_{l,o}  ) \leq 1 \\
&\quad\;\;  0\leq n_{l,o} \leq n_{max}
\end{align}
\end{subequations}

\begin{thm}
The problem \textbf{P2} is concave and the learner's optimal strategy is given by:
\begin{equation}
n^*_{l,o}=\left\{\begin{matrix}
 0& \zeta^2_{l,o}\tau_{l,o}-\zeta_{l,o}^1 < 0\\ 
\frac{ \zeta^2_{l,o} (1-G_o A^0_{l,o})}{G_o(\zeta_{l,o}^1 A^2_{l,o} + \zeta_{l,o}^2 A^1_{l,o})} &  n_{l,o} \in [n_{min},n_{max})\\
n_{max} & n_{l,o} \geq n_{max}
\end{matrix}\right.
\end{equation}
where:
\begin{equation}
n_{min} = \frac{\zeta^0_{l,o}}{\zeta^2_{l,o}\tau_{l,o}-\zeta_{l,o}^1 }
\end{equation}
\end{thm}

\textit{Proof.} It is readily obvious that the optimization problem \textbf{P2} is a Linear Program, and hence, it follows directly that it is concave. Moreover, the term $\zeta^2_{l,o}\tau_{l,o}-\zeta_{l,o}^1$ in the objective is the variable linear coefficient, and it can represent the net utility given the orchestrator's strategy. As a result, if the net utility is positive, the learner will try to maximize it by maximizing the amount of participation until the time frame period is finished. 
Conversely, if the net utility is negative, the learner will not participate in the learning process. The rest of the proof can be found in Appendix \ref{appA}.

\subsection{Second round: Deriving Orchestrators' Strategy}
Recall that the number of global cycles will be set according to the weakest learner in the group, given the fact that it utilizes its full-time period. Consequently, the constraint (14b) becomes equality, and the number of global cycles can be given by:
\begin{equation}
G_o = \max \left(G_{min}, \frac{1}{A^2_{\tilde{l},o} \tau_{l,o} n^*_{\tilde{l},o} + A^1_{\tilde{l},o} n^*_{\tilde{l},o} + A^0_{\tilde{l},o}} \right)
\end{equation}
where $\tilde{l}$ is the index of the weakest learner, and it can be determined as follows $\tilde{l} =  \arg \underset{l}{\min} ~ \frac{\tilde{f}_{l}}{\tilde{d}_{l,o}}$, as the learner with the less CPU frequency and greater distance to its associated orchestrator causes more delay in the learning process. 

Afterward, we derive the orchestrator's strategy set for the service monetary price. To ensure learners participation, the net utility term from (15) has to be positive, and it follows that:
\begin{equation}
\tau_{l,o} \geq \frac{\zeta^1_{l,o}}{\left (\frac{C_o \rho_{l,o}}{R_{max}} - \frac{\mu C_o f_l}{E_{max}} \right)}
\end{equation}
Since we know that $\tau_{min} \leq \tau_{l,o} \leq \tau_{max}$, we can thus define the lower and upper bounds for the monetary price as follows:
\begin{equation}
\underline{\rho_{l,o}} = \frac{\zeta^1_{l,o}}{\tau_{max}} - \frac{R_{max} \mu f_l}{E_{max}}
\end{equation}
\begin{equation}
\overline{\rho_{l,o}} = \frac{\zeta^1_{l,o}}{\tau_{min}} - \frac{R_{max} \mu f_l}{E_{max}}
\end{equation}

According to the above analysis, the orchestrator, which is the leader in the second round of the Stackelberg game, knows that there exists a Nash equilibrium among learners given any monetary service price and the number of local iterations within its strategy set. Therefore, by considering the learners' participation from (15) as a function of the service price, i.e., $n^*_{l,o}(\rho_{l,o})$, and setting the number of global cycles according to (17), the orchestrator can maximize its utility and determines its strategy by solving the following optimization:
\begin{subequations}\label{eqn:opt_orch}
\begin{align}
&\mathbf{P3}: \; \underset{\tau_{l,o}, \rho_{l,o}}{\max.}\; \frac{-1}{F_{max}|\mathcal{L}_o|} \sum_l \frac{c1}{G_o \tau^{c2}_{l,o}} - \frac{ G_o C_o}{P_{max}} \sum_l \rho_{l,o} \tau_{l,o} n^*_{l,o} \\
&s.t.\;\;\; \tau_{min} \leq \tau_{l,o} \leq \tau_{max} \\
&\quad\;\;  \underline{\rho_{l,o}}  \leq \rho_{l,o} \leq \overline{\rho_{l,o}} 
\end{align}
\end{subequations}
where $P_{max}$ is the maximum possible payment and $F_{max}$ is the maximum loss value. 
\begin{thm}
For $G_o C_o  
\rho_{l,o} \tau_{l,o} n_{l,o}\geq 2$, the problem \textbf{P3} is concave, and hence, there exists a Nash equilibrium for the second round of the game $(\tau^*_{l,o}, \rho^*_{l,o}, n^*_{l,o})$, where $(\tau^*_{l,o}, \rho^*_{l,o})$ is the maximizer of the orchestrator's utility.
\end{thm}

\textit{Proof.} The proof can be found in Appendix \ref{appB}.

It can be noticed that the term in the condition represents the participation revenue of the learner, and is verified to hold practically and shown in the result. Nevertheless, the optimization problem \textbf{P3} has no closed-form solution for $\tau^*_{l,o}$ and $\rho^*_{l,o}$. Hence, the orchestrator can find the optimal strategy using any optimization technique (e.g., gradient ascent, interior point method...etc.).
\begin{figure}
    \centering
    \includegraphics[scale=0.7]{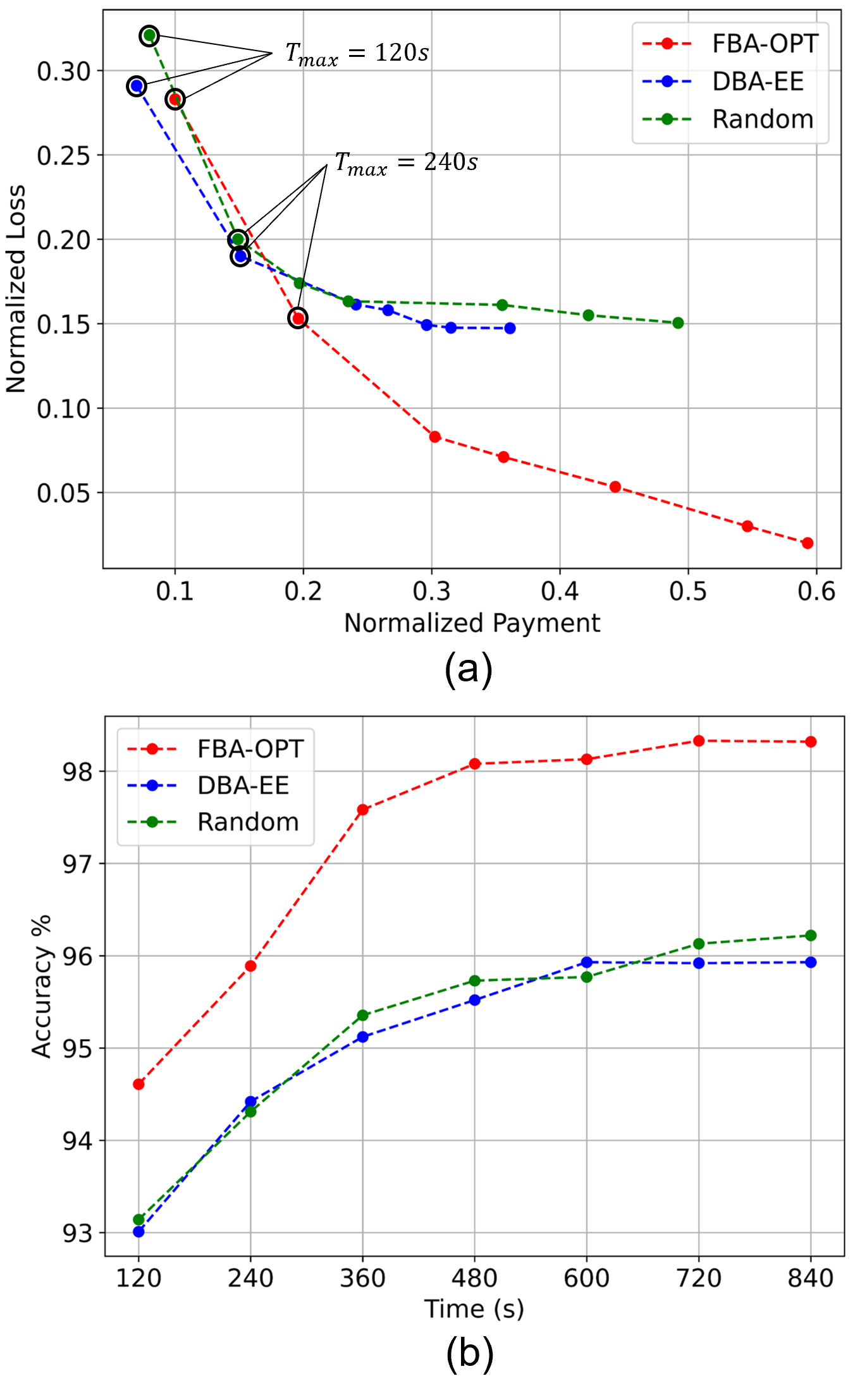}
    \caption{Orchestrators' performance comparison in terms of (a) utility trade-off (b) learning accuracy}
    \label{fig:orch}
\end{figure}
\section{Simulation Results}\label{sec:sim}
\subsection{Environment Setup}
We have conducted the experiments with 3 orchestrators and 50 learners. The learners' distances to the orchestrators were distributed uniformly randomly in the range of [5-50]m, and each learner can have one of the following processor frequencies [2.4,1.4,1.0] GHz. We have utilized the MNIST dataset for all the learning tasks. The learning models are multi-layer perceptrons (MLP) with [3,4,5] hidden layers, where higher number of layers represents higher model complexity. In the results, we show the average performance, e.g., the average learner utility and the average energy consumption.
\begin{figure*}
    \centering
    \includegraphics[scale=0.48]{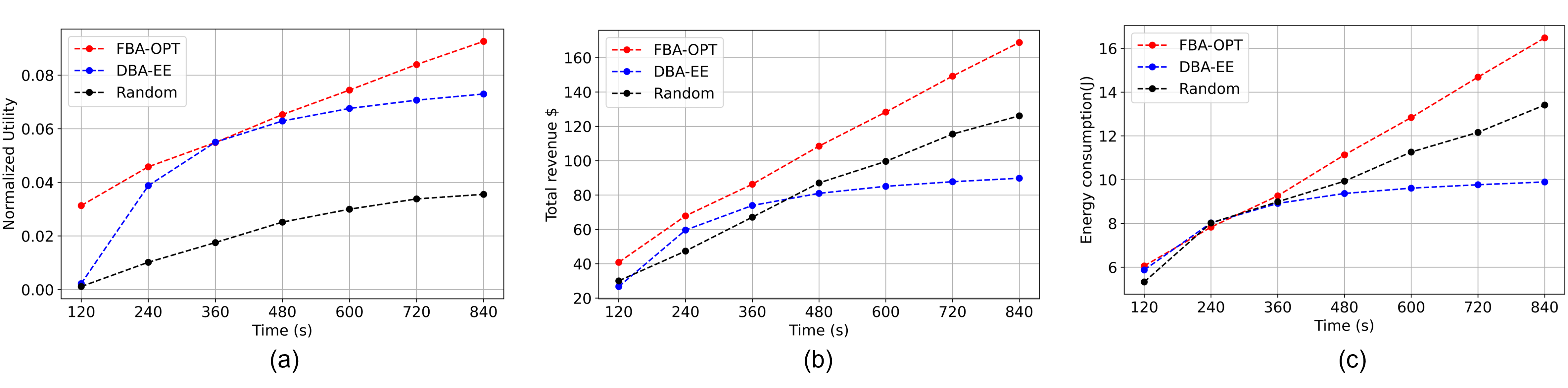}
    \caption{Learners' performance comparison in terms of (a) utility (b) total revenue (c) energy consumption}
    \label{fig:learner}
\end{figure*}
\subsection{Performance comparison}
We compare our approach which employs factor-based association and the optimal strategy (FBA-OPT) with the random strategy, and an energy-efficient technique that minimizes the energy consumption by considering the minimum possible number of iterations and global cycles, and employs a distance-based association, which we will refer to (DBA-EE). We compare the performance with respect to different time constraints, i.e., $T_{max}$.
The orchestrators' performance in terms of utility trade-off and learning accuracy is depicted in Fig. \ref{fig:orch}. In Fig. \ref{fig:orch} (a), we observe the trade-off curves for the utility function between the payment and the learning loss. Each point on the curves represents a different $T_{max}$. We can notice that, as we increase the time constraint, the total payment increases for all the approaches with a slight overpayment for our approach. However, the learning loss for the heuristic DBA-EE and the random strategies improve but then saturate with no improvements, while our approach keeps minimizing the learning loss. In Fig. \ref{fig:orch} (b), the learning accuracy of the tasks is shown. Our approach achieves higher accuracy and outperforms the other approaches for all the time constraints since it minimized the learning loss much more than the other approaches. 

The learners' performance is depicted in Fig \ref{fig:learner}. In Fig \ref{fig:learner} (a), the normalized utility is shown. It can be seen that for all approaches, the utility increases as $T_{max}$ increases. In fact, with more available time, the learner can increase the amount of participation to maximize its utility. However, our approach grants the best utility for the learners, while the random strategy is the worst. The heuristic DBA-EE approach at first starts increasing the utility, but then it starts to saturate. In fact, it limits the amount of participation in order to preserve its energy. The total revenue is shown in Fig \ref{fig:learner} (b). It can be noticed that our approach attains the best revenue for the learners, while the random approach attains better than the heuristic, since the latter limits its participation to preserve more energy. Lastly, the energy consumption is presented in  Fig \ref{fig:learner} (c). Similarly, the energy consumption increases as $T_{max}$ increases. However, the heuristic DBA-EE outperforms the other approaches and preserves energy the most, while our approach is opportunistic in terms of the revenue and consumes energy the most.

\section{Conclusion}\label{sec:conc}
In this work, we proposed an incentive mechanism for multi-orchestrator MEL by employing a 2-round Stackelberg game approach. In the first round, we employed a heuristic algorithm for the learners-orchestrators association. Whereas in the second round, we proved the existence of the Nash equilibrium in the game and derived a learner's optimal policy based on the associated orchestrator's incentive. Finally, numerical experiments have been conducted to show the performance of the proposed mechanism while being compared to other heuristic techniques.

\section*{Acknowledgement}
This work was made possible by NPRP grant \# NPRP12S-0305-190231 from the Qatar National Research Fund (a member of Qatar Foundation). We also acknowledge the support of the Natural Sciences and Engineering Research Council of Canada (NSERC), [RGPIN-2020-06919]. The findings achieved herein are solely the responsibility of the authors.  

\renewcommand*{\bibfont}{\small}
\printbibliography

@article{edge_surveillance,
  title={Distributed deep learning model for intelligent video surveillance systems with edge computing},
  author={Chen, Jianguo and Li, Kenli and Deng, Qingying and Li, Keqin and Philip, S Yu},
  journal={IEEE Transactions on Industrial Informatics},
  year={2019},
  publisher={IEEE}
}

@article{edge_crowd,
  title={Robust mobile crowd sensing: When deep learning meets edge computing},
  author={Zhou, Zhenyu and Liao, Haijun and Gu, Bo and Huq, Kazi Mohammed Saidul and Mumtaz, Shahid and Rodriguez, Jonathan},
  journal={IEEE network},
  volume={32},
  number={4},
  pages={54--60},
  year={2018},
  publisher={IEEE}
}

@article{edge2017survey,
  title={A survey on mobile edge computing: The communication perspective},
  author={Mao, Yuyi and You, Changsheng and Zhang, Jun and Huang, Kaibin and Letaief, Khaled B},
  journal={IEEE Communications Surveys \& Tutorials},
  volume={19},
  number={4},
  pages={2322--2358},
  year={2017},
  publisher={IEEE}
}

@article{gunduz2020communicate,
  title={Communicate to learn at the edge},
  author={G{\"u}nd{\"u}z, Deniz and Kurka, David Burth and Jankowski, Mikolaj and Amiri, Mohammad Mohammadi and Ozfatura, Emre and Sreekumar, Sreejith},
  journal={IEEE Communications Magazine},
  volume={58},
  number={12},
  pages={14--19},
  year={2020},
  publisher={IEEE}
}

@inproceedings{ME_1st,
  title={Adaptive task allocation for mobile edge learning},
  author={Mohammad, Umair and Sorour, Sameh},
  booktitle={2019 IEEE Wireless Communications and Networking Conference Workshop (WCNCW)},
  pages={1--6},
  year={2019},
  organization={IEEE}
}

@inproceedings{mcmahan2017communication,
  title={Communication-efficient learning of deep networks from decentralized data},
  author={McMahan, Brendan and Moore, Eider and Ramage, Daniel and Hampson, Seth and y Arcas, Blaise Aguera},
  booktitle={Artificial intelligence and statistics},
  pages={1273--1282},
  year={2017},
  organization={PMLR}
}

@inproceedings{edge_control,
  title={When edge meets learning: Adaptive control for resource-constrained distributed machine learning},
  author={Wang, Shiqiang and Tuor, Tiffany and Salonidis, Theodoros and Leung, Kin K and Makaya, Christian and He, Ting and Chan, Kevin},
  booktitle={IEEE INFOCOM 2018-IEEE Conference on Computer Communications},
  pages={63--71},
  year={2018},
  organization={IEEE}
}

@article{zhan2020learning_incentive,
  title={A learning-based incentive mechanism for federated learning},
  author={Zhan, Yufeng and Li, Peng and Qu, Zhihao and Zeng, Deze and Guo, Song},
  journal={IEEE Internet of Things Journal},
  volume={7},
  number={7},
  pages={6360--6368},
  year={2020},
  publisher={IEEE}
}

@article{sarikaya2019motivating_SB1,
  title={Motivating workers in federated learning: A stackelberg game perspective},
  author={Sarikaya, Yunus and Ercetin, Ozgur},
  journal={IEEE Networking Letters},
  volume={2},
  number={1},
  pages={23--27},
  year={2019},
  publisher={IEEE}
}

@inproceedings{zeng2020fmore_auction1,
  title={Fmore: An incentive scheme of multi-dimensional auction for federated learning in mec},
  author={Zeng, Rongfei and Zhang, Shixun and Wang, Jiaqi and Chu, Xiaowen},
  booktitle={2020 IEEE 40th International Conference on Distributed Computing Systems (ICDCS)},
  pages={278--288},
  year={2020},
  organization={IEEE}
}

@article{le2021incentive_auction2,
  title={An incentive mechanism for federated learning in wireless cellular network: An auction approach},
  author={Le, Tra Huong Thi and Tran, Nguyen H and Tun, Yan Kyaw and Nguyen, Minh NH and Pandey, Shashi Raj and Han, Zhu and Hong, Choong Seon},
  journal={IEEE Transactions on Wireless Communications},
  year={2021},
  publisher={IEEE}
}

@article{allahham2021energy,
  title={Energy-Efficient Multi-Orchestrator Mobile Edge Learning},
  author={Allahham, Mhd Saria and Sorour, Sameh and Mohamed, Amr and Erbad, Aiman and Guizani, Mohsen},
  journal={arXiv preprint arXiv:2109.00757},
  year={2021}
}

@article{mao2017survey_edge_energy,
  title={A survey on mobile edge computing: The communication perspective},
  author={Mao, Yuyi and You, Changsheng and Zhang, Jun and Huang, Kaibin and Letaief, Khaled B},
  journal={IEEE Communications Surveys \& Tutorials},
  volume={19},
  number={4},
  pages={2322--2358},
  year={2017},
  publisher={IEEE}
}

@book{myerson2013game,
  title={Game theory},
  author={Myerson, Roger B},
  year={2013},
  publisher={Harvard university press}
}

@article{cai2020d2d_straggler,
  title={D2D-enabled data sharing for distributed machine learning at wireless network edge},
  author={Cai, Xiaoran and Mo, Xiaopeng and Chen, Junyang and Xu, Jie},
  journal={IEEE Wireless Communications Letters},
  volume={9},
  number={9},
  pages={1457--1461},
  year={2020},
  publisher={IEEE}
}

@article{mohammad2020task_asynchronous,
  title={Task allocation for asynchronous mobile edge learning with delay and energy constraints},
  author={Mohammad, Umair and Sorour, Sameh and Hefeida, Mohamed},
  journal={arXiv preprint arXiv:2012.00143},
  year={2020}
}

\appendices

\section{}
\label{appA}
It is readily obvious that the optimization problem \textbf{P2} is a Linear Program, and hence, it follows directly that it is concave. Moreover, the term $\zeta^2_{l,o}\tau_{l,o}-\zeta_{l,o}^1$ in the objective is the variable linear coefficient, and it can represent the net utility given the orchestrator's strategy. As a result, if the net utility is positive, the learner will try to maximize it by maximizing the amount of participation until the time frame period is finished. Conversely, if the net utility is negative, the learner will not participate in the learning process. Thus, we can have constraint (14b) as equality in case of participation, and with some rearrangements we have the following:
\begin{equation}
\tau_{l,o}  = \frac{1- G_o( A^1_{l,o} n_{l,o} + A^0_{l,o} )}{n_{l,o} G_o A^2_{l,o} }
\end{equation}
However, we know from that in order to participate the following condition $\tau_o \geq \frac{\zeta^1_{l,o}}{\zeta^2_{l,o}} $ must hold. Therein, we have:
\begin{equation}
\frac{1- G_o( A^1_{l,o} n_{l,o} + A^0_{l,o} )}{n_{l,o} G_o A^2_{l,o} } \geq \frac{\zeta^1_{l,o}}{\zeta^2_{l,o}}
\end{equation}
and by rearranging the terms, we can have the following upper bound on the amount of participation:
\begin{equation}
n_{l,o} \leq \frac{ \zeta^2_{l,o} (1-G_o A^0_{l,o})}{G_o(\zeta_{l,o}^1 A^2_{l,o} + \zeta_{l,o}^2 A^1_{l,o})}
\end{equation}
Finally, since we know the utility function (14a) for the learner has to be positive, we can have the following lower bound:
\begin{equation}
n_{l,o} \geq \frac{\zeta^0_{l,o}}{\zeta^2_{l,o}\tau_{l,o}-\zeta_{l,o}^1 }
\end{equation}
If the amount of participation is less than the lower bound, it will result in a negative utility, hence the learner will not participate. On the other hand, the learner cannot participate with more than $n_{max}$, which is the maximum amount specified by the orchestrator. \hfill $\square$

\section{}
\label{appB}
In order to prove the concavity of the problem \textbf{P3}, we have to show that the Hessian matrix of the utility function is negative semi-definite. In practice, we can be set set the parameter $c2$ to 1 as in \cite{allahham2021energy}, and assume $P_{max} = 1$ and $\frac{c1}{F_{max} |\mathcal{L}_o|} = 1$. For the ease of reading, we will present the variables without the subscripts, i.e., $\tau_{l,o}$ as $\tau$ and $n_{l,o}$ as $n$...etc. Moreover, we present $\frac{d n_{l,o}}{d \rho_{l,o}}$ as $n'$, and to avoid confusions we introduce the following variables:
\begin{description}
        \item a = $A^2$, b = $A^1$, c = $A^0$
        \item A = $\zeta^2$, B = $\zeta^1$
        \item $\alpha = \frac{C_o}{R_{max}}$, $\beta = \frac{\mu C_o f_l}{E_{max}}~$  such that: $\zeta^2 = \alpha \rho - \beta $
\end{description}

Therein, we can have the following first derivatives:
\begin{equation}
\frac{\partial U}{\partial \tau} = - G C \rho n + \frac{1}{\tau^2 G}
\end{equation}
\begin{equation}
\frac{\partial U}{\partial \rho} = - G C \tau (\rho n' + n)
\end{equation}
where:
\begin{equation}
n = \frac{ A (1-Gc)}{G(aB + bA)}
\end{equation}
\begin{equation}
n' = \frac{B a\alpha (1-Gc)}{G(aB + bA)^2}
\end{equation}
Afterward, we can have the second derivatives as follows:
\begin{equation}
\frac{\partial^2 U}{\partial \tau^2} = - \frac{2}{\tau^3 G}
\end{equation}
\begin{equation}
\frac{\partial^2 U}{\partial \rho^2} = - G C \tau (2 n' + \rho n'')
\end{equation}

\begin{equation}
\frac{\partial U}{\partial \tau \rho} = \frac{\partial U}{\partial \rho  \tau} = - G C (\rho n' + n)
\end{equation}
where:
\begin{equation}
n'' = \frac{-2a\alpha^2bB (1-Gc)}{G(aB + bA)^3}
\end{equation}
We can notice that $n'\geq 0$ and $n'' \leq 0$. After that, we can express the Hessian matrix as follows:
\begin{equation}
\mathbf{H} = \begin{bmatrix}
- \frac{2}{\tau^3 G} & - G C (\rho n' + n)\\ 
- G C (\rho n' + n) &  - G C \tau (2 n' + \rho n'')
\end{bmatrix}
\end{equation}
For \textbf{H} to be negative semi-definite, its determinant and sub-determinant have to be negative, where the sub-determinant $- \frac{2}{\tau^3 G}$ is obviously negative. Therein:
\begin{equation}
\begin{split}
det(\mathbf{H}) = det \begin{vmatrix}
- \frac{2}{\tau^3 G} & - G C (\rho n' + n)\\ 
- G C (\rho n' + n) &  - G C \tau (2 n' + \rho n'')
\end{vmatrix}\\ = \frac{2C}{\tau^2}(2 n' + \rho n'') - G^2 C^2 (\rho n' + n)^2 \underset{?}{\leq} 0 \\
\frac{2C}{\tau^2}(2 n' + \rho n'') \underset{?}{\leq}  G^2 C^2 (\rho n' + n)^2 \\
2(2 n' + \rho n'') \underset{?}{\leq}  \tau^2 G^2 C \left ( (\rho n')^2+ 2\rho n' n + n^2 \right )\\
4 n'  \underset{?}{\leq}  \tau^2 G^2 C \left ( (\rho n')^2+ 2\rho n' n + n^2 \right ) - 2 \rho n''
\end{split}
\end{equation}
Since $n'\geq 0$, $n''\leq 0$ and $\tau,G \geq 1$, it is sufficient to show the following:
\begin{equation}
4 n'  \leq  \tau^2 G^2 C  2\rho n' n 
\end{equation}
or simply:
\begin{equation}
\tau G C \rho n \geq 2
\end{equation}
which completes the proof. \hfill $\square$
\end{document}